# Capacity Bounds for Dirty Paper with Exponential Dirt

M. Monemizadeh, S. Hajizadeh, G. A. Hodtani, and S. A. Seyedin
Department of Electrical Engineering
Ferdowsi University of Mashhad
Mashhad, Iran
mostafamonemizadeh@gmail.com, saeed.hajizadeh1367@gmail.com, ghodtani@gmail.com, and seyedin@um.ac.ir

*Abstract*—The additive exponential noise channel with additive exponential interference (AENC-AEI) known non-causally at the transmitter is studied. This channel can be considered as an exponential version of the discrete memoryless channel with state known non-causally at the encoder considered by Gelfand and Pinsker. We make use of Gelf'and-Pinsker's classic capacity Theorem to derive inner and outer bounds on the capacity of this channel under a non-negative input constraint as well as a constraint on the mean value of the input. First we obtain an outer bound for AENC-AEI. Then by using the input distribution achieving the outer bound, we derive an inner bound which this inner bound coincides with the obtained outer bound at high signal to noise ratios (SNRs) and therefore, gives the capacity of the AENC-AEI at high SNRs.

## I. INTRODUCTION

The capacity of a single-user channel with additive exponential noise (AEN), in which input is assumed to be a non-negative random variable independent of the noise and subject to a mean value constraint was determined by Verdu in [1]. Verdu showed that (i) the capacity of such channel is identical to the capacity of its complex-valued additive white Gaussian noise (AWGN) counterpart, and (ii) the optimal distribution achieving the capacity is not exponential, but it is a mixed distribution consisting of an exponential distribution and a point mass. In spite of a number of notable similarities between AEN channel and AWGN channel [1]-[2], there are basic and intrinsic differences between them due to a few reasons: the first of which is that the addition of two independent Gaussian random variables (RVs) is a Gaussian RV with a small alteration in its mean and variance values while the story is much different as it comes to the addition of two independent exponential RVs. The addition of two independent exponential RVs not only is not an exponential RV, but also the achieved distribution does not have a closed-form differential entropy, especially when the two added RVs have different mean values. The second reason is that in contrary to a Gaussian RV, an exponential RV is distributed only on the right hand side of the Euclidean two-dimensional space and therefore we cannot enjoy the flexibilities that a symmetric system brings into hand. Notice that the AEN channel is of practical importance. It naturally arises in optical and non-coherent communication systems. Exponential noise can also model phase noises and phase interferences in phase modulation schemes.

Discrete memoryless channels (DMCs) with side information (SI) were first studied by Shannon [3] where he found the capacity of a single-user DMC with SI causally available at the transmitter. Gelf'and and Pinsker [4] characterized the capacity of a single-user DMC when SI is non-causally available at the transmitter. Costa [5] described the Gaussian version of [4] in his celebrated and whimsically titled paper and showed that the capacity of AWGN channel when also afflicted by additive white Gaussian interference, is equal to the capacity of the AWGN channel as far as the additive interference is non-causally known to the sender so that it can adapt its signal to totally eliminate the negative effect of the interference.

In this paper we introduce the AEN channel afflicted by an extra i.i.d. additive interference sequence exponentially distributed and independent of the noise of the channel (an exponential version of [4]).

In this paper, we will make use of Gelf'and-Pinsker's capacity characterization to derive general inner and outer bounds on the capacity of this channel when the input is assumed to be a non-negative real number and is also subject to a mean value constraint. First we obtain an outer bound for this channel. Then by using the input distribution achieving outer bound, we derive an inner bound which coincides with the outer bound at high signal to noise ratio (SNR) and hence, gives the capacity of the channel at high SNRs. The rest of the paper is organized as follows. In Section II, we define the AEN channel with additive exponentially distributed interference (AEI) known non-causally at the transmitter (AENC-AEI). All the main results are presented in Section III. In Section IV, some numerical results are provided to evaluate the proposed lower and upper bounds.

## II. CHANNEL MODEL AND BACKGROUND

Fix positive scalars $m_x$, $m_s$ and $m_z$. Let $S$ and $Z$ be two independent exponentially distributed random variables with means $m_s$ and $m_z$, respectively, and $X$ be a non-negative random variable independent of $S$ and $Z$ and subject to input mean constraint $m_x$. Also, let $\bar{X}$ be a non-negative random variable

independent of $S$ and $Z$ and with mean $m_x$ and a mixed distribution

$$f_{\bar{X}}(x) = \frac{m_z}{m_x + m_z}\delta(x) + \frac{m_x}{(m_x + m_z)^2}e^{\frac{-x}{m_x+m_z}}u(x). \quad (1)$$

where $\delta(x)$ and $u(x)$ are Dirac delta and unit step functions, respectively, and defined as

$$\delta(x) = \begin{cases} 1 & x = 0, \\ 0 & \text{otherwis,} \end{cases} \quad (2)$$

$$u(x) = \begin{cases} 1 & x > 0, \\ 0 & \text{otherwise.} \end{cases} \quad (3)$$

The Laplace transform of the mixed probability density function (pdf) of the random variable $\bar{X}$ is

$$\mathcal{L}\{f_{\bar{X}}(x)\} = \int_0^\infty e^{-sx} f_{\bar{X}}(x) dx = \frac{1 + m_z s}{1 + (m_x + m_z)s}. \quad (4)$$

It is worth noting that the random variable $T = \bar{X} + Z$ is exponentially distributed with mean $m_t = m_x + m_z$. Throughout this paper the auxiliary random variable $U$ is defined in the simple form $U = X + S$. Moreover, let $\bar{U} = \bar{X} + S, Y = X + S + Z$ and $\bar{Y} = \bar{X} + S + Z$.

Consider the communication channel depicted in Fig. 1. which is a single-user AEN channel with additive, exponentially distributed interference (AEI) known non-causally at the transmitter. This channel, denoted by AENC-AEI, has the following characteristics.

D1. Channel is an additive memoryless channel with discrete time and continuous alphabets in which the channel output is given by $Y_i = X_i + S_i + Z_i$, for $i = 1, ..., n$.
D2. $Z^n$ is an independent identically distributed (i.i.d.) sequence with mean $m_z$ and exponential distribution.
D3. $S^n$ is an i.i.d. sequence with mean $m_s$ and exponential distribution which is known non-causally to the transmitter and is independent of noise $Z$.
D4. Channel input $X$ is a non-negative random variable independent of $S$ and $Z$ and subject to an input mean constraint, i.e. we have $\frac{1}{n}\sum_{i=1}^n X_i \leq m_x$.

Gelf'and and Pinsker [4] showed that the capacity of a single-user DMC when side information sequence $S^n$ is non-causally available at the transmitter is given by

$$C = \max_{p(u,x|s)} \{I(U;Y) - I(U;S)\}, \quad (5)$$

where the maximum is over all joint distributions $p(s,u,x,y)$ that factor as $p(s)p(u,x|s)p(y|x,s)$ and $U$ is an auxiliary random variable.

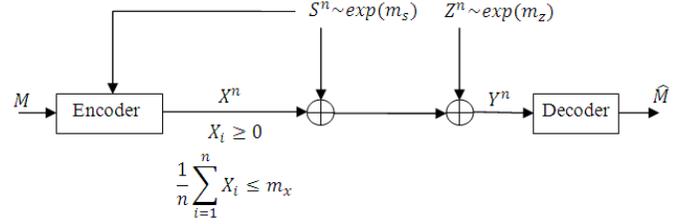

Figure 1. The additive exponential noise channel with additive exponential interference known non-causally at the transmitter.

Costa [5] obtained a Gaussian version of Gelf'and-Pinsker's capacity Theorem and showed that the capacity of AWGN channel with an input power constraint when also afflicted by additive white Gaussian interference, is equal to the capacity of the AWGN channel as long as full knowledge of this additive interference is given to the encoder so that it can adapt its signal to totally eliminate the negative effect of the interference.

### III. MAIN RESULTS: INNER AND OUTER BOUNDS ON CAPACITY FOR THE AENC-AEI

In this section we provide the capacity bounds for the AENC-AEI depicted in Fig.1. and defined with properties D1-D4. We first obtain an outer bound as follows. Throughout this paper $H(\cdot)$ denotes the differential entropy.

#### A. An Outer Bound:

We can obtain easily an outer bound as below.

$$C = \max_{p(u,x|s)} \{I(U;Y) - I(U;S)\}$$

$$= \max_{p(u,x|s)} \{-H(U|Y) + H(U|S)\} \quad (6)$$

$$\leq \max_{p(u,x|s)} \{-H(U|Y,S) + H(U|S)\} \quad (7)$$

$$= \max_{p(u,x|s)} I(U;Y|S) \quad (8)$$

$$= \max_{p(u,x|s)} \{H(Y|S) - H(Y|U,S)\} \quad (9)$$

$$= \max_{p(x|s)} \{H(X + Z) - H(Z)\} \quad (10)$$

$$= H(\bar{X} + Z) - H(Z) \quad (11)$$

$$= \log(e(m_x + m_z)) - \log(em_z) \quad (12)$$

$$= \log\left(1 + \frac{m_x}{m_z}\right) \quad (13)$$

$$\triangleq C_{out} \quad (14)$$

where (7) follows from that fact that conditioning reduces entropy and (11) is due to the random variable $\bar{X}$ is such that the random variable $T = \bar{X} + Z$ is exponentially distributed. It is worth noting that the exponential distribution has maximum differential entropy for a given mean constraint mentioned in D4.

Now, using random variable $\bar{X}$, we obtain an inner bound for the capacity of AENC-AEI. This inner bound coincides with the outer bound (14) at high SNRs and hence, gives the capacity of AENC-AEI at high SNRs.

## B. An Inner Bound:

Considering (1) and (5) we can write:

$$C = \max_{p(u,x|s)} \{I(U;Y) - I(U;S)\}$$

$$\geq \max_{p^*(u,x|s)} \{I(U;Y) - I(U;S)\} \quad (15)$$

$$= \max_{p^*(u,x|s)} \{H(Y) - H(Y|U) - H(U) + H(U|S)\} \quad (16)$$

$$= H(\bar{X} + S + Z) - H(Z) - H(\bar{X} + S) + H(\bar{X}) \quad (17)$$

$$\triangleq C_{in} \quad (18)$$

where $p^*(u,x|s)$ is a subset of the set of all distributions $p(u,x|s)$ in which $X = \bar{X}$ and thereupon $U = \bar{U}$. We now calculate each term on the right side of (17) separately. As we know the random variable $Z$ has an exponential pdf:

$$f_Z(z) = \frac{1}{m_z} e^{\frac{-z}{m_z}} u(z), \quad (19)$$

therefore,

$$H(Z) = -\int f_Z(z) \log f_Z(z) dz \quad (20)$$

$$= \log(e m_z) \quad (21)$$

Considering that the random variable $\bar{X}$ has a mixed distribution (1), $H(\bar{X})$ is

$$H(\bar{X}) = -\int f_{\bar{X}}(x) \log f_{\bar{X}}(x) dx \quad (22)$$

$$= -\frac{m_z}{m_x + m_z} \log\left(\frac{m_z}{m_x + m_z}\right) -$$

$$\int_{x>0} \left(\frac{m_x}{(m_x + m_z)^2} e^{\frac{-x}{m_x + m_z}}\right) \log\left(\frac{m_x}{(m_x + m_z)^2} e^{\frac{-x}{m_x + m_z}}\right) dx \quad (23)$$

$$= -\frac{m_z}{m_x + m_z} \log\left(\frac{m_z}{m_x + m_z}\right)$$

$$+ \frac{m_x}{m_x + m_z} \log\left(\frac{e(m_x + m_z)^2}{m_x}\right). \quad (24)$$

To calculate $H(\bar{Y})$ we need to find the pdf of the random variable $\bar{Y} = \bar{X} + S + Z$. The Laplace transform of the exponential pdf of the random variable $T = \bar{X} + Z$ is

$$\mathcal{L}\left\{f_{\bar{X}+Z}(t) = \frac{1}{m_x + m_z} e^{\frac{-t}{m_x + m_z}} u(t)\right\} = \frac{1}{1 + (m_x + m_z)s} \quad (25)$$

therefore, the Laplace transform of the pdf of the random variable $\bar{Y} = \bar{X} + S + Z$ is

$$\mathcal{L}\{f_{\bar{Y}}(y)\} = \frac{1}{1 + (m_x + m_z)s} \times \frac{1}{1 + (m_s)s}$$

$$= \frac{1}{m_x + m_z - m_s}\left(\frac{m_x + m_z}{1 + (m_x + m_z)s} - \frac{m_s}{1 + (m_s)s}\right) \quad (26)$$

By applying the inverse Laplace transform to (26), we obtain $f_{\bar{Y}}(y)$.

$$f_{\bar{Y}}(y) = \frac{1}{m_x + m_z - m_s}\left(e^{\frac{-y}{m_x+m_z}} - e^{\frac{-y}{m_s}}\right) u(y), \quad (27)$$

consequently,

$$H(\bar{Y}) = -\int f_{\bar{Y}}(y) \log f_{\bar{Y}}(y) dy \quad (28)$$

$$= -\int_{y>0} \left(\frac{e^{\frac{-y}{m_x+m_z}} - e^{\frac{-y}{m_s}}}{m_x + m_z - m_s}\right) \log\left(\frac{e^{\frac{-y}{m_x+m_z}} - e^{\frac{-y}{m_s}}}{m_x + m_z - m_s}\right) dy. \quad (29)$$

Before evaluating the integral of (29), we need to express the Taylor series of the $\ln(1 + x)$ around $x = 0$ (Maclaurin series).

*Remark 1:* The Taylor series of the $\ln(1 + x)$ around $x = 0$ (Maclaurin series) is:

$$\ln(1+x) = \sum_{k=1}^{\infty} \frac{(-1)^{k+1}}{k} x^k, \quad -1 < x \leq 1 \quad (30)$$

Note that the series approximation converges to the function only in the region $-1 < x \leq 1$ and is only valid within this range.

The term $\ln\left(e^{\frac{-y}{m_x+m_z}} - e^{\frac{-y}{m_s}}\right)$ in (29) can be written as $\ln\left(e^{\frac{-y}{m_x+m_z}}(1 - e^{-yA_1})\right)$ or $\ln\left(-e^{\frac{-y}{m_s}}(1 - e^{-yA_2})\right)$ in which $A_1 = \frac{1}{m_s} - \frac{1}{m_x+m_z}$ and $A_2 = -A_1 = \frac{1}{m_x+m_z} - \frac{1}{m_s}$. So, considering the convergence region of $\ln(1 + x)$ and also knowing that $y > 0$, we have two cases for (29). For the series convergence in the first case, it should be $A_1 > 0$ or equivalently $m_x + m_z > m_s$. Therefore, we have:

$$H(\bar{Y}) =$$

$$-\int_{y>0} \left(\frac{e^{\frac{-y}{m_x+m_z}} - e^{\frac{-y}{m_s}}}{m_x + m_z - m_s}\right) \log\left(\frac{e^{\frac{-y}{m_x+m_z}}(1 - e^{-yA_1})}{m_x + m_z - m_s}\right) dy \quad (31)$$

$$= -\int_{y>0} \left(\frac{e^{\frac{-y}{m_x+m_z}} - e^{\frac{-y}{m_s}}}{m_x + m_z - m_s}\right) \log\left(\frac{e^{\frac{-y}{m_x+m_z}}}{m_x + m_z - m_s}\right) dy$$

$$-\log e \int_{y>0} \left(\frac{e^{\frac{-y}{m_x+m_z}} - e^{\frac{-y}{m_s}}}{m_x + m_z - m_s}\right) \ln(1 - e^{-yA_1}) dy \quad (32)$$

$$= \log(m_x + m_z - m_s) + \left(\frac{m_x + m_z + m_s}{m_x + m_z}\right) \log e$$

$$-\log e \int_{y>0} \left(\frac{e^{\frac{-y}{m_x+m_z}} - e^{\frac{-y}{m_s}}}{m_x + m_z - m_s}\right) \ln(1 - e^{-yA_1}) dy \quad (33)$$

Also, using (30) we can write:

$$\ln(1 - e^{-yA_1}) = -\sum_{k=1}^{\infty} \frac{e^{-kyA_1}}{k}, \quad y > 0, A_1 > 0 \tag{34}$$

consequently,

$$\int_{y>0} \left(e^{\frac{-y}{m_x+m_z}} - e^{\frac{-y}{m_s}}\right) \ln(1 - e^{-yA_1}) dy$$

$$= -\int_{y>0} \left(e^{\frac{-y}{m_x+m_z}} - e^{\frac{-y}{m_s}}\right) \sum_{k=1}^{\infty} \frac{e^{-kyA_1}}{k} dy \tag{35}$$

$$= \sum_{k=1}^{\infty} \frac{1}{k} \left\{ \int_{y>0} \left(e^{-y\left(\frac{k+1}{m_s} - \frac{k}{m_x+m_z}\right)} - e^{-y\left(\frac{k}{m_s} - \frac{k-1}{m_x+m_z}\right)}\right) dy \right\} \tag{36}$$

$$= \sum_{k=1}^{\infty} \frac{1}{k} \left\{ \frac{1}{\frac{k+1}{m_s} - \frac{k}{m_x+m_z}} - \frac{1}{\frac{k}{m_s} - \frac{k-1}{m_x+m_z}} \right\} \tag{37}$$

$$= \sum_{k=1}^{\infty} \frac{1}{k} \left\{ \frac{1}{F(k+1)} - \frac{1}{F(k)} \right\} \tag{38}$$

where $F(k) = \frac{k}{m_s} - \frac{k-1}{m_x+m_z}$. Therefore, for $A_1 > 0$ or equivalently $m_x + m_z > m_s$, $H(\bar{Y})$ is

$$H(\bar{Y}) = \log(m_x + m_z - m_s) + \left(\frac{m_x + m_z + m_s}{m_x + m_z}\right) \log e$$
$$+ \left(\frac{\log e}{m_x + m_z - m_s}\right) \left(\sum_{k=1}^{\infty} \frac{1}{k} \left\{\frac{1}{F(k)} - \frac{1}{F(k+1)}\right\}\right) \tag{39}$$

Similarly, for $A_2 = -A_1 > 0$ or equivalently $m_x + m_z < m_s$, $H(\bar{Y})$ is

$$H(\bar{Y}) = \log(m_s - m_x - m_z) + \left(\frac{m_x + m_z + m_s}{m_s}\right) \log e$$
$$- \left(\frac{\log e}{m_x + m_z - m_s}\right) \left(\sum_{k=1}^{\infty} \frac{1}{k} \left\{\frac{1}{G(k)} - \frac{1}{G(k+1)}\right\}\right) \tag{40}$$

where $G(k) = \frac{k}{m_x+m_z} - \frac{k-1}{m_s}$.

Finally, we calculate $H(\bar{U})$. So, first we find the pdf of the random variable $\bar{U} = \bar{X} + S$. Using Laplace transform we have:

$$\mathcal{L}\{f_{\bar{U}}(u)\} = \mathcal{L}\{f_{S^*}(s)\} \times \mathcal{L}\{f_{\bar{X}}(x)\} \tag{41}$$

$$= \frac{1}{1 + m_s s} \times \frac{1 + m_z s}{1 + (m_x + m_z)s} \tag{42}$$

$$= \frac{1}{m_x + m_z - m_s} \left(\frac{m_x}{1 + (m_x + m_z)s} - \frac{m_s - m_z}{1 + m_s s}\right) \tag{43}$$

Therefore, by using the inverse Laplace transform, we obtain the pdf of the random variable $\bar{U}$ as

$$f_{\bar{U}}(u) =$$
$$\frac{1}{m_x + m_z - m_s} \left(\frac{m_x}{m_x + m_z} e^{\frac{-u}{m_x+m_z}} - \frac{m_s - m_z}{m_s} e^{\frac{-u}{m_s}}\right) u(u) \tag{44}$$

Consequently, we have

$$H(\bar{U}) = -\int f_{\bar{U}}(u) \log f_{\bar{U}}(u) du \tag{45}$$

$$= -\int_{u>0} \left\{ \left(\frac{\frac{m_x}{m_x+m_z} e^{\frac{-u}{m_x+m_z}} - \frac{m_s-m_z}{m_s} e^{\frac{-u}{m_s}}}{m_x + m_z - m_s}\right) \times \right.$$
$$\left. \log\left(\frac{\frac{m_x}{m_x+m_z} e^{\frac{-u}{m_x+m_z}} - \frac{m_s-m_z}{m_s} e^{\frac{-u}{m_s}}}{m_x + m_z - m_s}\right) du \right\} \tag{46}$$

Similar to the expansion of (29), the term $\ln\left(\frac{m_x}{m_x+m_z} e^{\frac{-u}{m_x+m_z}} - \frac{m_s-m_z}{m_s} e^{\frac{-u}{m_s}}\right)$ in (46) can be written as $\ln\left(\left(\frac{m_x}{m_x+m_z} e^{\frac{-u}{m_x+m_z}}\right)(1 - B_1 e^{-uA_1})\right)$ or $\ln\left(\left(-\frac{m_s-m_z}{m_s} e^{\frac{-u}{m_s}}\right)(1 - B_2 e^{-uA_2})\right)$ in which $A_1 = \frac{1}{m_s} - \frac{1}{m_x+m_z}$, $B_1 = \frac{(m_s-m_z)(m_x+m_z)}{m_x m_s}$ and $A_2 = -A_1 = \frac{1}{m_x+m_z} - \frac{1}{m_s}$, $B_2 = \frac{1}{B_1} = \frac{m_x m_s}{(m_s-m_z)(m_x+m_z)}$. So, considering the convergence region of $\ln(1+x)$ and also knowing that $u > 0$, we have two cases for (46). For the series convergence in the first case it should be $A_1 > 0$ and $-1 \leq B_1 < 1$. Accordingly, we can write

$$H(\bar{U}) = -\int_{u>0} \left\{ \left(\frac{\frac{m_x}{m_x+m_z} e^{\frac{-u}{m_x+m_z}} - \frac{m_s-m_z}{m_s} e^{\frac{-u}{m_s}}}{m_x + m_z - m_s}\right) \times \right.$$
$$\left. \log\left(\frac{\left(\frac{m_x}{m_x+m_z} e^{\frac{-u}{m_x+m_z}}\right)(1 - B_1 e^{-uA_1})}{m_x + m_z - m_s}\right) du \right\} \tag{47}$$

$$= -\int_{u>0} \left\{ \left(\frac{\frac{m_x}{m_x+m_z} e^{\frac{-u}{m_x+m_z}} - \frac{m_s-m_z}{m_s} e^{\frac{-u}{m_s}}}{m_x + m_z - m_s}\right) \times \right.$$
$$\left. \log\left(\frac{\frac{m_x}{m_x+m_z} e^{\frac{-u}{m_x+m_z}}}{m_x + m_z - m_s}\right) du \right\} - (\log e) \times$$
$$\int_{u>0} \left(\frac{\frac{m_x}{m_x+m_z} e^{\frac{-u}{m_x+m_z}} - \frac{m_s-m_z}{m_s} e^{\frac{-u}{m_s}}}{m_x + m_z - m_s}\right) \ln(1 - B_1 e^{-uA_1}) du$$

$$= \log\left(\frac{m_x + m_z - m_s}{\frac{m_x}{m_x + m_z}}\right) + \left(\frac{\log e}{m_x + m_z - m_s}\right) \times \left(m_x - \frac{m_s(m_s - m_z)}{m_x + m_z}\right) - (\log e) \times$$

$$\int_{u>0} \left(\frac{\frac{m_x}{m_x+m_z}e^{\frac{-u}{m_x+m_z}} - \frac{m_s-m_z}{m_s}e^{\frac{-u}{m_s}}}{m_x + m_z - m_s}\right) \ln(1 - B_1 e^{-uA_1}) du \quad (49)$$

By using

$$\ln(1 - B_1 e^{-uA_1}) = -\sum_{k=1}^{\infty} \frac{B_1^k e^{-kuA_1}}{k},$$
$$u > 0, A_1 > 0, -1 \le B_1 < 1 \quad (50)$$

we have

$$\int_{u>0} \left(\frac{m_x}{m_x+m_z}e^{\frac{-u}{m_x+m_z}} - \frac{m_s-m_z}{m_s}e^{\frac{-u}{m_s}}\right) \ln(1 - B_1 e^{-uA_1}) du$$

$$= -\int_{y>0} \left\{ \left(\frac{m_x}{m_x+m_z}e^{\frac{-u}{m_x+m_z}} - \frac{m_s-m_z}{m_s}e^{\frac{-u}{m_s}}\right) \times \right.$$
$$\left. \sum_{k=1}^{\infty} \frac{B_1^k e^{-kuA_1}}{k} du \right\} \quad (51)$$

$$= \sum_{k=1}^{\infty} \frac{B_1^k}{k} \left\{ \int_{u>0} \left[\frac{m_s - m_z}{m_s} e^{-u\left(\frac{k+1}{m_s} - \frac{k}{m_x+m_z}\right)} \right.\right.$$
$$\left.\left. - \frac{m_x}{m_x+m_z} e^{-u\left(\frac{k}{m_s} - \frac{k-1}{m_x+m_z}\right)}\right] du \right\} \quad (52)$$

$$= \sum_{k=1}^{\infty} \frac{B_1^k}{k} \left\{ \frac{\frac{m_s - m_z}{m_s}}{\frac{k+1}{m_s} - \frac{k}{m_x+m_z}} - \frac{\frac{m_x}{m_x+m_z}}{\frac{k}{m_s} - \frac{k-1}{m_x+m_z}} \right\} \quad (53)$$

$$= \sum_{k=1}^{\infty} \frac{B_1^k}{k} \left\{ \frac{\frac{m_s - m_z}{m_s}}{F(k+1)} - \frac{\frac{m_x}{m_x+m_z}}{F(k)} \right\} \quad (54)$$

where $F(k) = \frac{k}{m_s} - \frac{k-1}{m_x+m_z}$. Hence, for $A_1 > 0$, $-1 \le B_1 < 1$, $H(\bar{U})$ is

$$H(\bar{U}) = \log\left(\frac{m_x + m_z - m_s}{\frac{m_x}{m_x + m_z}}\right) + \left(\frac{\log e}{m_x + m_z - m_s}\right)\left(m_x - \frac{m_s(m_s - m_z)}{m_x + m_z}\right) + \left(\frac{\log e}{m_x + m_z - m_s}\right)\left(\sum_{k=1}^{\infty} \frac{B_1^k}{k} \left\{\frac{\frac{m_x}{m_x+m_z}}{F(k)} - \frac{\frac{m_s-m_z}{m_s}}{F(k+1)}\right\}\right) \quad (55)$$

Similarly, for $A_2 = -A_1 > 0$ and $-1 \le B_2 = \frac{1}{B_1} < 1$, $H(\bar{U})$ is

$$H(\bar{U}) = \log\left(\frac{m_s - m_x - m_z}{\frac{m_s - m_z}{m_s}}\right) + \left(\frac{\log e}{m_x + m_z - m_s}\right)\left(\frac{m_x(m_x + m_z)}{m_s} - (m_s - m_z)\right) - \left(\frac{\log e}{m_x + m_z - m_s}\right)\left(\sum_{k=1}^{\infty} \frac{B_2^k}{k}\left\{\frac{\frac{m_s - m_z}{m_s}}{G(k)} - \frac{\frac{m_x}{m_x + m_z}}{G(k+1)}\right\}\right) \quad (56)$$

where $G(k) = \frac{k}{m_x+m_z} - \frac{k-1}{m_s}$. Therefore, from (18), (21), (24), (39), (55) and after simplification, the inner bound of the capacity for $A_1 > 0$ and $-1 \le B_1 < 1$ is obtained as shown in (57). Also, from (18), (21), (24), (40) and (56), the inner bound of the capacity for $A_2 = -A_1 > 0$ and $-1 \le B_2 = \frac{1}{B_1} < 1$ is as shown in (58) at the bottom of the page.

---

$$C \ge C_{in}^1$$
$$= \frac{m_z}{m_x + m_z}\log\left(\frac{m_x}{m_z^2}\right) + \frac{m_x}{m_x + m_z}\log\left(\frac{m_x + m_z}{m_z}\right) + \left(\frac{\log e}{m_x + m_z - m_s}\right)\left(\sum_{k=1}^{\infty} \frac{1}{k}\left\{\frac{1 - B_1^k \frac{m_x}{m_x+m_z}}{F(k)} - \frac{1 - B_1^k \frac{m_s - m_z}{m_s}}{F(k+1)}\right\}\right) \quad (57)$$

---

$$C \ge C_{in}^2$$
$$= \frac{m_x}{m_x + m_z}\log\left(\frac{m_x + m_z}{m_x}\right) + \left(\frac{\log e}{m_s}\right)(m_z) + \log\left(\frac{m_s - m_z}{m_s}\right) + \log\left(\frac{m_x + m_z}{m_z}\right) + \frac{m_z}{m_x + m_z}\log\left(\frac{1}{em_z}\right)$$
$$+ \left(\frac{\log e}{m_s - m_x - m_z}\right)\left(\sum_{k=1}^{\infty} \frac{1}{k}\left\{\frac{1 - B_2^k \frac{m_s - m_z}{m_s}}{G(k)} - \frac{1 - B_2^k \frac{m_x}{m_x+m_z}}{G(k+1)}\right\}\right) \quad (58)$$

*Corollary 1:* By considering (14), (57) for $m_x \gg m_z \cong m_s$ we can readily obtain

$$C \cong C_{in} \cong C_{out} \cong \log\left(\frac{m_x}{m_z}\right) = \log(SNR) \quad (59)$$

in other words, for high SNRs the inner and outer bounds are tight and give the capacity of AENC-AEI. Note that in this case by considering (1), the optimal input distribution is an exponential distribution with mean $m_x + m_z \cong m_x$.

## IV. NUMERICAL RESULTS

In this section, we compute the lower and upper bounds of the capacity of AENC-AEI which we have derived in previous section. Fig.2. depicts the capacity bounds and capacity gap of AENC-AEI for $m_s \geq m_z$ and several values of $m_z$: $10$, $10^2$, $10^4$, and $10^6$. It is seen that the inner bound coincides with the outer bound at high SNR and hence, gives the capacity of AENC-AEI at high SNR which is equal to $C = \log\left(\frac{m_x}{m_z}\right) = \log(SNR)$.

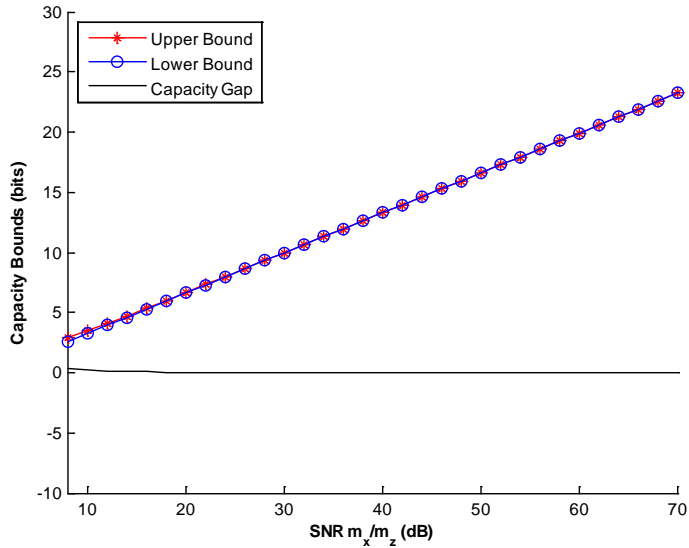

(a) $m_z = 10$

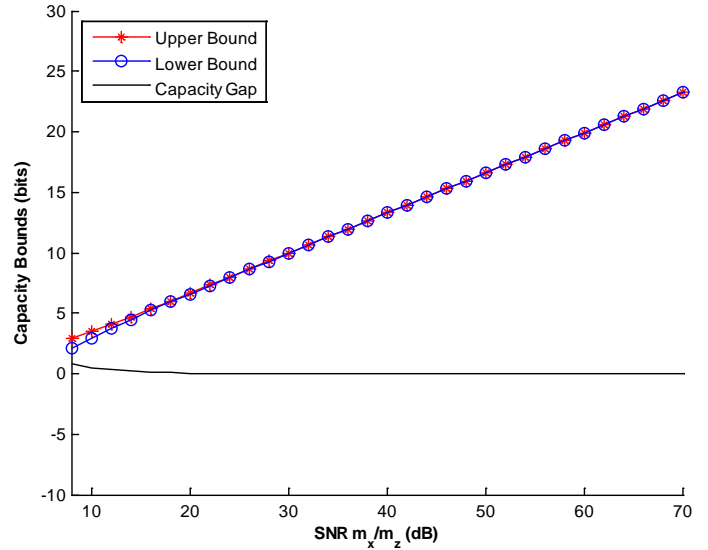

(b) $m_z = 10^2$

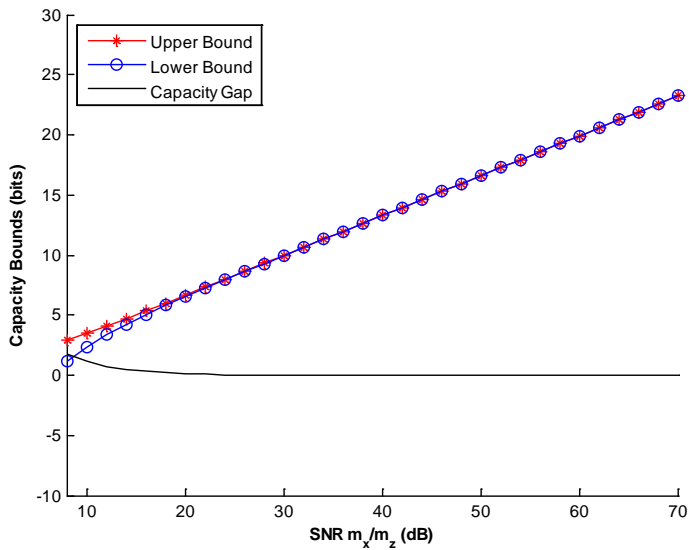

(c) $m_z = 10^4$

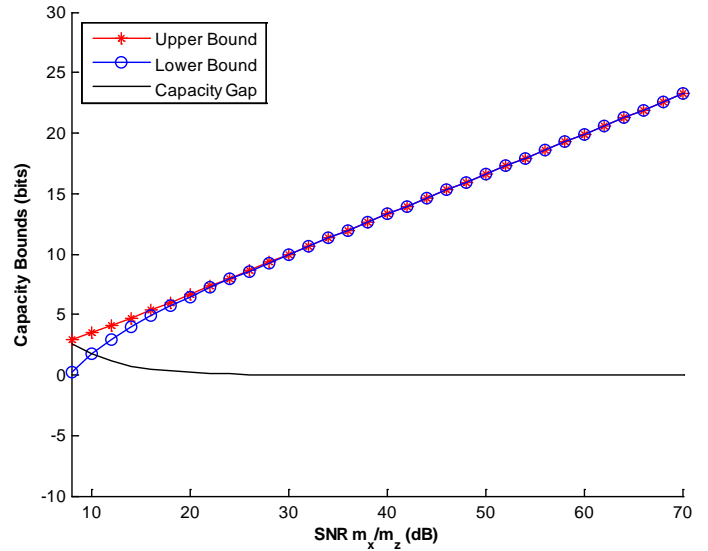

(d) $m_z = 10^6$

Figure. 2. Capacity bounds and capacity gap of the AENC-AEI for $m_s \geq m_z$ and several values of $m_z$.

## REFERENCES

[1] S. Verdu, "The exponential distribution in information theory," *Prob. Inf Transm.*, vol. 32, no. 1, pp. 86–95, Jan-Mar 1996.

[2] A. Martinez, "Communication by energy modulation: The additive exponential noise channel," *IEEE Trans. Inform. Theory,* vol. 57, pp. 3333-3351, June 2011.

[3] C. E. Shannon, "Channels with side information at the transmitter," *IBM Journal of Research and Development,* vol. 2, no. 4, pp. 289–293, October 1958.

[4] S. Gel'fand and M. Pinsker, "Coding for channels with random parameters," *Probl. Contr. and Inf. Theory,* vol. 9, no. 1, pp. 19–31, 1980.

[5] M. H. M. Costa, "Writing on dirty paper," *IEEE Trans. Inf. Theory,* vol. 29, no. 3, pp. 439–441, May 1983.